\documentclass[a4paper,10pt]{article}
\usepackage{epsfig,amsmath}
\usepackage{color}

\begin{document}

\newcommand{\highlight}{\color{blue}}

\title{Comment on "Pinched Flow Fractionation: Continuous Size Separation of Particles Utilizing a Laminar Flow Profile in a Pinched Microchannel"}

\author{Niels Asger Mortensen\footnote{email: nam@mic.dtu.dk}\\MIC -- Department of Micro and
Nanotechnology, NanoDTU\\ Technical University of Denmark,\\
  DTU - Building 345east, DK-2800 Kongens Lyngby, Denmark
}

\date{\today}

\maketitle

In a recent paper Yamada {\it et al.}~\cite{Yamada:2004a} propose
the novel concept of {\it "pinched flow fractionation"} (PFF) for
the continuous size separation and analysis of particles in
microfabricated lab-on-a-chip devices. In their description of the
basic principle they claim that especially the width of the
pinched and broadened segments will affect the separation
performance. In the following we comment on the physics behind
this statement.

Considering the steady-state laminar flow of an incompressible
Newtonian liquid, the flow field $\boldsymbol v$ is governed by
Stokes equation
\begin{equation}\label{eq:Stokes}
0 = -{\boldsymbol \nabla} p + \eta \nabla^2{\boldsymbol v},
\end{equation}
where $\eta$ is the viscosity and $p$ the pressure. We for
simplicity assume that we have a flow rate $Q$ and a
well-developed Poiseuille flow with a parabolic flow profile in
both the pinched segment (of width $w_P$) as well as in the
broadened segment (of width $w_B$) at the detection line. Next,
consider a streamline at a distance $y_P$ from the, say, lower
wall of the channel in the pinched region. In the broadened
segment, we consider the same streamline and denote the distance
to the channel wall by $y_B$. From simple conservation of mass and
the scale-invariance of the linear Stokes equation,
Eq.~(\ref{eq:Stokes}), we quite intuitively arrive at
\begin{eqnarray}
\frac{y_B}{w_B}=\frac{y_P}{w_P}\Longleftrightarrow y_B=
y_P\frac{w_B}{w_P}
\end{eqnarray}
thus demonstrating the foundation of the quite simple geometrical
scaling. In the context of PFF Eq.~(2) is often referred to as the
{\it assumption of linear amplification}! Considering a particle
of diameter $D$ located in the upper part of the channel we have
$y_P=w_P-D/2$ whereby we arrive at Eq.~(Y1). In the above analysis
we have made no assumptions about the detailed flow pattern in the
transition between the pinched and the broadened segments and thus
the result is completely general/universal, i.e. different device
geometries with creeping flow will obey the same scaling law. To
further support this, Fig.~\ref{fig2} shows finite-element
simulations of the Stokes equation for two cases with different
geometrical transitions between the pinched and the broadened
segments. In full agreement with the above discussion we see that
the stream lines are universal at the detection line while the two
families of stream lines do of course deviate from each other in
the transition region where the two geometries differ. In the
relevant limit where convection dominates over diffusion (high
P\'eclet number) the particles dynamics is governed by Stokes drag
where the drag force is in the direction of the velocity
field~\cite{Bruus:2007} so that the streamlines become
trajectories for the particles. At the detection line, the
separation of Stokes-driven particles is thus independent of the
detailed geometry. This strongly contrasts the claim by Yamada
{\it et al.}~\cite{Yamada:2004a} in their description of the basic
principle: {\it "Also, microchannel geometries (especially the
width of the pinched and broadened segments) would affect the
separation performance, since particle movement is dominated by
flow profiles inside the channel"}~\cite{Yamada:2004a}.
Interestingly, their claim seems well supported by their
experimental data while it is in conflict with Eq.~(Y1) [or
equivalently Eq.~(2) above]. This discrepancy with simple theory
suggests more complicated particle dynamics beyond simple Stokes
drag with particle trajectories deviating from the streamline
picture, especially in the vicinity of channel walls. Indeed, the
simple Stokes drag picture should be modified in the vicinity of
channel walls (Fax{\'e}n correction) where the flow velocity is
indeed highly non-uniform on the length scale of the particle
diameter. This combined with e.g. surface roughness could
introduce more complicated particle rotations and interactions
which might tend to reflect the detailed geometry, thus giving
sorting dynamics beyond the bounds of Eq.~(2). In conclusion,
engineering of PFF devices will have to face modelling of detailed
particle dynamics beyond simple creeping-flow streamline
considerations.

\vspace{1cm}

{\it Acknowledgments.} I would like to thank H. Bruus for useful
discussions and K. M{\o}lhave, A.V. Larsen, and A. Kristensen for
directing my attention to the present problem and the paper by
Yamada {\it et al.}.

%\bibliographystyle{jps}
%\bibliography{pinched}

\newpage

\begin{figure}[t!]
\begin{center}
\epsfig{file=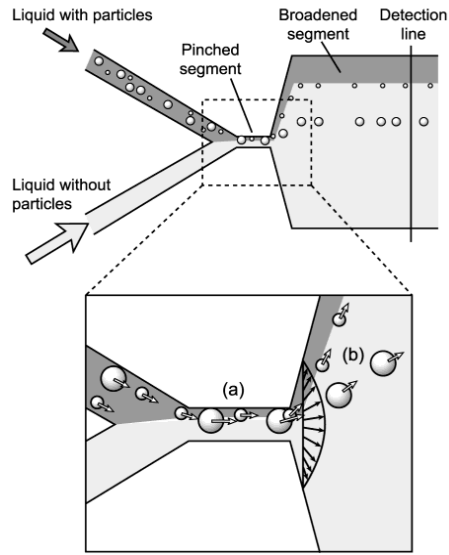,width=0.5\columnwidth,clip}
\end{center}
\caption{Principle of pinched flow fractionation (reproduced from
Ref.~\cite{Yamada:2004a}). } \label{fig1}
\end{figure}

\begin{figure}[b!]
\begin{center}
\epsfig{file=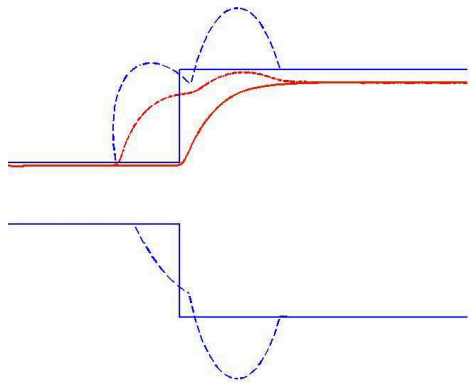,width=0.5\columnwidth,clip}
\end{center}
\caption{Stream lines obtained by finite-element simulation of
Stokes equation, Eq.~(\ref{eq:Stokes}), for an abrupt change in
geometry (solid red line) and a more arbitrary geometry (dashed
red line). The corresponding geometries are indicated by the
super-imposed blue lines. } \label{fig2}
\end{figure}

\end{document}